\documentclass[10pt,letterpaper]{article}

\usepackage[left=1in,right=1in,top=1in,bottom=1in]{geometry}
\usepackage[T1]{fontenc}
\usepackage[utf8]{inputenc} 
\usepackage{lmodern}
\usepackage{microtype}
\usepackage{url}

\usepackage{amsmath,amssymb}
\usepackage{graphicx}
\usepackage{cite}
\usepackage{indentfirst}
\usepackage{authblk}
\makeatletter
\renewcommand\AB@affilnote[1]{\raisebox{0.35ex}{\scriptsize#1}\hspace{0.08em}}
\usepackage[font=small,labelfont=bf]{caption}
\usepackage[hidelinks]{hyperref}
\usepackage[ruled,vlined,linesnumbered]{algorithm2e}
\usepackage{titlesec}
\usepackage{enumitem}

\usepackage[table,xcdraw,dvipsnames]{xcolor}
\usepackage{booktabs}
\usepackage{multirow}
\usepackage{pifont}
\usepackage{textcomp}

\definecolor{npsgreen}{HTML}{E1F1EC}
\definecolor{npsred}{HTML}{FEECDD}
\definecolor{cvprblue}{HTML}{DEEBF7}
\definecolor{icmlred}{HTML}{E58579}
\definecolor{cvprgreen}{HTML}{BBDBA6}
\definecolor{cvprorange}{HTML}{F4B183}
\definecolor{lightblue}{HTML}{DCEAF7}
\definecolor{lightgreen}{HTML}{E8F3E1}
\definecolor{lightorange}{HTML}{F9E7D2}
\definecolor{graybg}{gray}{0.95}

\newcommand{\std}[1]{\textcolor{gray}{\scriptsize{$\pm$#1}}}
\newcommand{\gain}[1]{\textcolor{ForestGreen}{\scriptsize\textbf{(+#1)}}}
\newcommand{\loss}[1]{\textcolor{RoyalBlue}{\scriptsize\textbf{(-#1)}}}

\usepackage{tikz}
\usetikzlibrary{shapes.geometric, arrows.meta, positioning, fit, backgrounds, shadows, calc}

\tikzset{
    basebox/.style={rectangle, draw=gray!60, thick, rounded corners=3pt, align=center, fill=white, font=\small, drop shadow={opacity=0.05}},
    cloudbox/.style={basebox, fill=cvprblue!50, minimum height=1cm},
    edgebox/.style={basebox, fill=npsgreen!80, minimum height=1cm},
    uebox/.style={basebox, fill=gray!10, minimum height=1cm},
    groupbox/.style={draw=#1, dashed, thick, rounded corners=5pt, inner sep=12pt},
    arrow/.style={->, >=Stealth, thick, draw=gray!80},
    dasharrow/.style={->, >=Stealth, dashed, thick, draw=gray!80},
    decision/.style={diamond, aspect=1.8, draw=gray!60, thick, fill=gray!10, align=center, font=\scriptsize\bfseries, inner sep=0pt, drop shadow={opacity=0.05}}
}

\renewcommand{\thesection}{\Roman{section}}
\renewcommand{\thesubsection}{\Alph{subsection}}
\titleformat{\section}{\normalfont\bfseries\centering}{\thesection.}{0.6em}{\MakeUppercase}
\titleformat{name=\section,numberless}{\normalfont\bfseries\centering}{}{0pt}{\MakeUppercase}
\titleformat{\subsection}{\normalfont\bfseries}{\thesubsection.}{0.6em}{}
\titlespacing*{\section}{0pt}{1.25\baselineskip}{0.75\baselineskip}
\titlespacing*{\subsection}{0pt}{0.9\baselineskip}{0.45\baselineskip}
\setlist[enumerate]{leftmargin=1.5em,itemsep=0.25em,topsep=0.35em}

\title{Edge-Side Residual Timing and Frequency Control\\
for Software-Defined Ground Stations in 5G NTN Uplinks}

\author{}
\date{} 
\newcommand{\affline}[2]{%
  \noindent\hangindent=1.5em\hangafter=1%
  \makebox[1.5em][r]{\textsuperscript{#1}}#2\par
}

\begin{document}
\maketitle
\vspace{-4.4em}
\begin{center}
{\fontsize{12}{15}\selectfont
Longji He$^{1,2}$, Elena Emma Wang$^{3}$, Xichun Wang$^{4}$, Juntao Xu$^{1}$, and Jiaming Li$^{*4}$}
\end{center}
\vspace{0.55em}
{\fontsize{10}{14}\selectfont
\affline{1}{OgCloud Limited, Guangzhou, GD, China, \texttt{keith@ogcloud.com}}
\affline{2}{College of Engineering, The Pennsylvania State University, University Park, PA, USA}
\affline{3}{The Meadows School, \texttt{elena.wang.mathlete@gmail.com}}
\affline{4}{School of Informatics, Computing \& Cyber Systems, Steve Sanghi College of Engineering,\\
Northern Arizona University, Flagstaff, AZ, USA}
}
\vspace{0.85em}

\begin{abstract}

This paper studies a ground-segment implementation problem in 5G non-terrestrial networks (NTN): once UE-side geometric pre-compensation has produced a coarse timing/frequency prior, can an edge-side residual loop keep the uplink inside an NR-feasible operating region under rapid LEO dynamics? We examine this question with a software-defined ground station (SDGS) design that keeps the coarse prior at the UE and closes the residual timing-advance (TA) / carrier-frequency-offset (CFO) loop at the ground-station edge. This paper takes a systems-and-control view rather than proposing a full-stack intelligent architecture. Its evidence base consists of a March 2026 hardware-in-the-loop (HIL) campaign and a companion uncertainty analysis. The HIL campaign includes same-window reference runs collected on the same platform with edge residual control disabled, but it does not include a cloud-loop benchmark. The placement claim is therefore architectural and control-oriented rather than a head-to-head cloud-versus-edge proof. In the Shenzhen steady-state tracking interval, the edge-controlled mode lowers mean RTT from $70.51 \pm 2.34$ ms to $32.84 \pm 2.56$ ms and, within the retained Layer-3 transport mapping, improves artifact-level goodput from $80.14 \pm 0.14$ Mbps to $196.04 \pm 1.87$ Mbps relative to that reference configuration. Across four ground-station locations, the closed-loop controller keeps residual TA P95 at $0.49\,\mu\text{s}$ and residual CFO P95 within $76$--$77$ Hz. Together with the uncertainty analysis, these observations support a bounded claim: an edge-side residual timing/frequency loop can keep the SDGS uplink in a more stable NR-feasible operating regime under the assumptions retained in the current HIL artifact.

\end{abstract}

\section{Introduction}

The integration of non-terrestrial networks (NTN) into 5G has moved beyond a theoretical proposal and into an active standards and deployment path. 3GPP has already incorporated NR NTN procedures into the Release 17 family, and the resulting design space now includes practical questions that are no longer purely waveform-theoretic: what timing/frequency functions should remain at the UE, what should be absorbed by the network implementation, and how much residual control must be kept close to the ground segment \cite{3gpp_ts_38_214,3gpp_ts_38_211}. This matters because LEO service links combine fast geometry variation, repeated handover windows, and tight timing/frequency tolerances that become operationally visible at the ground segment.

Software-defined ground stations (SDGS) are a natural place to study that problem. Their value lies in virtualizing satellite-ground functions through NFV/SDN-style software stacks \cite{etsi_nfv_man,kratos_openspace,aws_ground_station} and exposing a control-placement decision that proprietary monolithic systems often hide. Recent public measurements of Starlink and other LEO services further emphasize that nominal downlink rates do not tell the whole story: latency engineering, packet-loss bursts, and route variability all shape whether a link is operationally usable \cite{starlink_latency_2024, starlink_specs_2025, starlink_aviation_2025, wirelessmoves_starlink_loss_2024, zenodo_starlink_ping_2025}.

In this setting, the paper focuses on an implementation gap rather than on a standards novelty claim. 3GPP NTN procedures clearly motivate UE-side timing/frequency pre-compensation, but they do not prescribe a single ground-segment realization for the residual loop under SDGS deployment constraints. Existing literature contains rich analyses of open-loop compensation, receiver tracking, and learning-based networking. What remains relatively under-specified is the narrow but important systems question addressed here: after the UE has formed a coarse geometric prior, can a latency-sensitive residual timing/frequency loop be run stably at the SDGS edge, and what kind of evidence can a Layer-3 HIL artifact provide about that design choice?

\subsection{Problem Statement}

For the uplink, the difficulty is not only that LEO channels are dynamic. The harder problem is that UE-side orbital prediction, ground-side residual tracking, and transport-layer link usability live on different timescales. A coarse geometric prior can be computed near the terminal, but the remaining TA/CFO error must still be corrected where observability and response time are both adequate. If that residual loop is pushed too far away from the ground segment, control latency grows exactly where Doppler slope, handover preparation, and queue evolution are changing most quickly. The result is a placement-sensitive control problem. In the current manuscript, however, we do not claim a universal cloud-versus-edge winner from direct A/B benchmarking; instead, we evaluate one concrete edge-side realization against same-campaign open-loop reference runs and use delay sensitivity plus system constraints to motivate the placement argument.

\subsection{Study Scope and Objectives}

The paper asks a narrower question than a full end-to-end NTN redesign: how much uplink stability can be recovered when coarse UE-side geometry is paired with a fast SDGS edge residual loop? We restrict attention to the uplink timing/frequency control path, the local adaptation logic that shares the same edge workflow, and whether the resulting residual TA/CFO remains in an NR-feasible regime. The retained evidence is correspondingly narrow and explicit: a March 2026 HIL campaign for transport-level behavior, same-campaign reference runs on the same platform with the residual loop disabled, and a separate uncertainty analysis for perturbation robustness.

\subsection{Contributions}

This paper contributes one SDGS control path that is explicit, bounded, and testable. First, we treat the NTN uplink problem as a timing-and-frequency question: the UE forms a coarse geometric prior and the SDGS refines the TA/CFO estimates. Second, we recast the timing/frequency claim as an error-budget problem tied to ephemeris, GNSS, clock, and propagation uncertainty rather than to ideal orbit knowledge. Third, we define a same-campaign reference configuration and a protocol-defined steady-state tracking interval, so that the HIL evidence is interpreted as a controlled comparison rather than as an unqualified benchmark claim. Fourth, we separate evidence layers explicitly: the HIL campaign supports a bounded transport-level systems claim under the retained Layer-3 mapping, while the uncertainty analysis and delay sweep bound TA/CFO feasibility and controller sensitivity.

\section{Related Work}

Work on software-defined ground infrastructure has mainly focused on virtualization and orchestration. Platforms such as Kratos OpenSpace and AWS Ground Station show that satellite-ground functions can be separated from monolithic hardware and moved into cloud-based systems \cite{kratos_openspace, aws_ground_station}. That line of work, however, says comparatively little about the fast uplink control loop that becomes critical when an SDGS must follow LEO dynamics in real time.

A separate literature addresses Doppler compensation and synchronization under direct-to-satellite or LEO conditions. Classical communication models explain why radial motion at these orbits produces large frequency offsets \cite{proakis2007digital}. More application-specific studies examine open-loop prediction from orbital state \cite{farhat2025doppler} or closed-loop receiver tracking \cite{farhangian2020multi}. Learning-based estimators have also been explored \cite{ngo2023deep}, but they are usually presented as signal-processing components rather than as pieces of a deployable SDGS workflow.

Satellite-edge studies provide the third thread. Recent work argues that latency-sensitive decisions should move closer to the ground segment \cite{wen2025satelight, wang2025deep}, while public Starlink measurements show that packet loss and latency can fluctuate in ways that nominal throughput figures do not capture \cite{wirelessmoves_starlink_loss_2024, zenodo_starlink_ping_2025}. Our paper uses these threads more narrowly than many broad intelligent-network architectures do. The contribution is the claim that UE-side prediction, SDGS-side residual correction, and local link adaptation should be analyzed as one ground-segment control path.

\subsection{Compatibility with 3GPP NR NTN Procedures}

Our framework is designed to coexist with 3GPP NR NTN procedures rather than replace them. The UE-side geometric stage provides an \emph{initial} open-loop TA/CFO prior that narrows the residual range before the signal reaches the ground segment. The residual timing and frequency loop is then refined at the SDGS edge using observables that are available on the receive path. This is aligned with the broader NR NTN logic in which pre-compensation of the service link is associated with the UE while some Doppler-management details remain a network-implementation matter. In the evaluation, we therefore interpret the architecture as a prediction--refinement pipeline and report residual TA/CFO relative to the CP/SCS-oriented feasibility discussion in Section~III-C rather than as a replacement for standard NR tracking.

\section{System Model and Problem Formulation}

\subsection{System Architecture and Channel Setting}

\medskip\noindent\textbf{\textit{System architecture.}}\ 
We consider a SDGS network serving a LEO constellation with $N_s$ satellites in multiple orbital planes. The operational architecture comprises two active control tiers plus an optional offline analysis path, later read in Fig.~\ref{fig:sdgs_framework}. The UE tier contains user terminals with varying computational capabilities, from resource-constrained IoT devices to high-performance terminals, and is responsible for geometric derivation together with preliminary signal processing. The edge tier consists of ground stations equipped with edge computing nodes that absorb the fast residual timing/frequency loop and the local adaptation logic required by the same uplink session. An additional offline analysis path is retained for log aggregation, replay, and threshold retuning, but it is not part of the latency-sensitive residual loop claimed in this paper.

\medskip\noindent\textbf{\textit{Channel setting.}}\ 
The uplink channel between user $u$ and satellite $s$ experiences both large-scale path loss and small-scale fading \cite{proakis2007digital}:

\begin{equation}
y_{u,s}(t) = h_{u,s}(t) \cdot x_{u,s}(t) + n_{u,s}(t)
\end{equation}

where $h_{u,s}(t)$ incorporates Doppler shift $f_d^{u,s}(t)$ and delay spread $\tau_{u,s}(t)$. The dominant time variation in the present problem is the Doppler drift induced by fast LEO motion, which we model from the relative velocity vector:

\begin{equation}
f_d^{u,s}(t) = \frac{f_c}{c} \cdot \frac{d}{dt} \|r_u(t) - r_s(t)\|
\end{equation}

where $f_c$ is the carrier frequency, $c$ is the speed of light, and $r_u(t)$, $r_s(t)$ are position vectors. The UE-side geometric derivation module computes expected Doppler shifts using satellite two-line elements (TLE) and SGP4 propagation.

\subsection{Problem Formulation}

\medskip\noindent\textbf{\textit{Problem formulation.}}\ 
The operational objective is to maximize usable uplink throughput while respecting latency, power, and computational constraints:

\begin{equation}
\begin{aligned}
\max_{\mathbf{x}, \mathbf{a}} \quad & \sum_{u \in \mathcal{U}} \sum_{s \in \mathcal{S}} R_{u,s}(t) \\
\text{s.t.} \quad & R_{u,s}(t) \leq R_{\max}^{u,s} \\
& \text{Latency}_{u,s}(t) \leq L_{\max} \\
& P_u(t) \leq P_{\max}^u \\
& \mathcal{C}_u^{\text{comp}}(t) \leq \mathcal{C}_{u,\max}^{\text{comp}}
\end{aligned}
\end{equation}

where $R_{u,s}(t)$ is achievable rate, $\text{Latency}_{u,s}(t)$ includes propagation and processing delay, $P_u(t)$ is transmission power, and $\mathcal{C}_u^{\text{comp}}(t)$ is computational load.

To keep the control and transport layers mathematically connected without overstating what the retained artifact proves, we model the HIL engine as a regime-switching transport system:
\begin{equation}
s(t)=
\begin{cases}
\text{nominal}, & |\Delta\tau(t)| \le \tau_{\mathrm{cp}} \ \text{and}\ |\Delta f(t)| \le f_{\mathrm{scs}}\\
\text{degraded}, & \text{otherwise},
\end{cases}
\end{equation}
\begin{equation}
R_{u,s}(t)=
\begin{cases}
R_{u,s}^{(0)}(t), & s(t)=\text{nominal}\\
R_{u,s}^{(\mathrm{deg})}(t), & s(t)=\text{degraded},
\end{cases}
\qquad
L_{u,s}(t)=
\begin{cases}
L_{u,s}^{(0)}(t), & s(t)=\text{nominal}\\
L_{u,s}^{(0)}(t)+L_{u,s}^{(\mathrm{retx})}(t), & s(t)=\text{degraded},
\end{cases}
\end{equation}
where $\tau_{\mathrm{cp}}$ and $f_{\mathrm{scs}}$ are the TA/CFO regime thresholds used by the current HIL engine. In a waveform-level receiver, the residual-to-performance bridge would be expressed through ISI/ICI, SINR, and BLER. In the Layer-3 setup, the bridge is implemented as a threshold-triggered penalty that, once residual TA/CFO leaves the target range, injects additional delay, jitter, loss, and reduced artifact-level goodput. This model is intentionally explicit about its status: it preserves the direction of the causal bridge while not pretending that the HIL testbed is itself a full PHY emulator.

\subsection{Residual Error Budget and NR Feasibility}

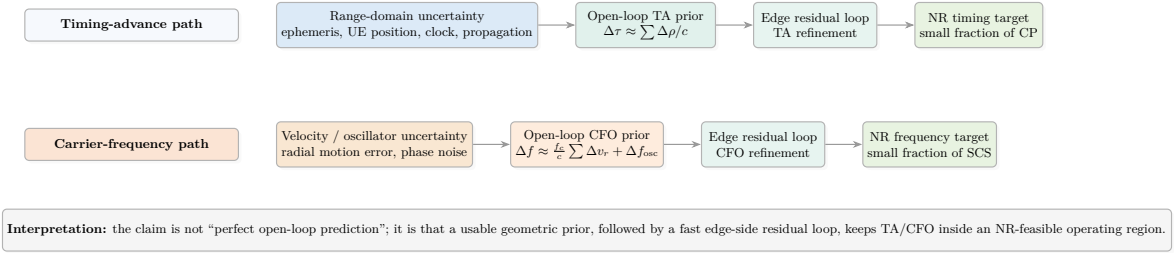
\begin{figure*}[t]
\centering
\resizebox{0.95\textwidth}{!}{
\begin{tikzpicture}[node distance=0.8cm and 1.0cm]
    \node[basebox, fill=cvprblue!28, minimum width=5.2cm, minimum height=0.78cm] (hh1) {\textbf{Timing-advance path}};
    \node[basebox, fill=cvprorange!35, minimum width=5.2cm, minimum height=0.78cm, below=2.1cm of hh1] (hh2) {\textbf{Carrier-frequency path}};

    \node[basebox, fill=lightblue, minimum width=4.2cm, minimum height=1.1cm, right=0.9cm of hh1] (t1) {Range-domain uncertainty\\ephemeris, UE position, clock, propagation};
    \node[basebox, fill=npsgreen, minimum width=3.4cm, minimum height=1.1cm, right=0.9cm of t1] (t2) {Open-loop TA prior\\$\Delta \tau \approx \sum \Delta \rho / c$};
    \node[edgebox, minimum width=3.0cm, minimum height=1.1cm, right=0.9cm of t2] (t3) {Edge residual loop\\TA refinement};
    \node[basebox, fill=lightgreen, minimum width=3.0cm, minimum height=1.1cm, right=0.9cm of t3] (t4) {NR timing target\\small fraction of CP};

    \node[basebox, fill=lightorange, minimum width=4.2cm, minimum height=1.1cm, right=0.9cm of hh2] (f1) {Velocity / oscillator uncertainty\\radial motion error, phase noise};
    \node[basebox, fill=npsred, minimum width=3.4cm, minimum height=1.1cm, right=0.9cm of f1] (f2) {Open-loop CFO prior\\$\Delta f \approx \frac{f_c}{c}\sum \Delta v_r + \Delta f_{\mathrm{osc}}$};
    \node[edgebox, minimum width=3.0cm, minimum height=1.1cm, right=0.9cm of f2] (f3) {Edge residual loop\\CFO refinement};
    \node[basebox, fill=lightgreen, minimum width=3.0cm, minimum height=1.1cm, right=0.9cm of f3] (f4) {NR frequency target\\small fraction of SCS};

    \draw[arrow] (t1.east) -- (t2.west);
    \draw[arrow] (t2.east) -- (t3.west);
    \draw[arrow] (t3.east) -- (t4.west);
    \draw[arrow] (f1.east) -- (f2.west);
    \draw[arrow] (f2.east) -- (f3.west);
    \draw[arrow] (f3.east) -- (f4.west);

    \node[basebox, fill=gray!8, minimum width=13.8cm, minimum height=1.02cm, below=1.0cm of f2] (note) {\textbf{Interpretation:} the claim is not ``perfect open-loop prediction''; it is that a usable geometric prior, followed by a fast edge-side residual loop, keeps TA/CFO inside an NR-feasible operating region.};
\end{tikzpicture}
}
\caption{Error-budget view of the TA/CFO control chain. Each lane should be read as a short sequence: uncertainty sources produce an open-loop prior, the SDGS edge loop tightens the residual, and the result is judged against a CP/SCS-oriented NR acceptance target.}
\label{fig:error_budget_chain}
\end{figure*}

\medskip\noindent\textbf{\textit{Residual error budget and NR feasibility.}}\ 
The proposed geometric pre-compensation provides an \emph{open-loop} prediction of timing advance (TA) and carrier-frequency offset (CFO) based on satellite ephemeris and UE position. In practice, the residual timing and frequency errors are dominated by imperfect ephemeris, UE positioning uncertainty, and UE/ground clock offsets. To make the timing claim self-consistent and aligned with NR procedures, we explicitly budget the dominant error sources and interpret the target relative to NR cyclic prefix (CP) and subcarrier spacing (SCS). Figure~\ref{fig:error_budget_chain} condenses this logic into a short error-budget chain. \emph{NR-oriented acceptance target.} Rather than claiming that raw ephemeris-only prediction universally achieves sub-$\mu$s timing accuracy, we target an \emph{NR-feasible} residual timing error after open-loop pre-compensation and closed-loop refinement. In NR uplink, the practical requirement is that the residual timing offset is small relative to the cyclic prefix (CP) for the selected numerology (SCS/CP), so that inter-symbol interference and inter-carrier interference remain bounded and the DMRS-based tracking loops remain effective. In this paper, we therefore report and interpret residual TA/CFO in a \emph{microsecond-level} regime (and corresponding residual CFO in the \emph{few-hundred-Hz to kHz} regime depending on SCS), and we explicitly evaluate robustness to ephemeris/position/clock perturbations (\ref{app:uncertainty}) to avoid over-claiming precision that is not supported by realistic TLE/SGP4 error characteristics.

\emph{Timing error model.} Let $\rho(t)$ be the true slant range and $\widehat{\rho}(t)$ the predicted range from UE-side geometry. The predicted one-way propagation delay is $\widehat{\tau}(t)=\widehat{\rho}(t)/c$, where $c$ is the speed of light. The residual timing error can be expressed as
\begin{equation}
\Delta \tau(t)=\tau(t)-\widehat{\tau}(t) \approx \frac{\Delta \rho_{\text{eph}}(t)+\Delta \rho_{\text{UE}}(t)+\Delta \rho_{\text{clk}}(t)+\Delta \rho_{\text{prop}}(t)}{c},
\end{equation}
where $\Delta \rho_{\text{eph}}$ captures ephemeris/orbit prediction error (e.g., TLE/SGP4 bias), $\Delta \rho_{\text{UE}}$ reflects UE position uncertainty (e.g., GNSS), $\Delta \rho_{\text{clk}}$ captures relative clock offset/drift mapped to an equivalent range error, and $\Delta \rho_{\text{prop}}$ represents unmodeled propagation contributions (e.g., ionospheric delay at S-band).

A practical design target is \emph{microsecond-level} open-loop timing (i.e., $|\Delta \tau|$ on the order of $\mathcal{O}(1\,\mu\text{s})$), which corresponds to an aggregate range error on the order of a few hundred meters ($1\,\mu\text{s}\approx 300\,\text{m}$). This level is achievable when (i) ephemeris errors are bounded through periodic updates and/or calibration at the ground segment, and (ii) UE position/clock uncertainty is managed (e.g., GNSS-aided timing and oscillator discipline). Importantly, NR timing is not purely open-loop: the network can refine TA via standard closed-loop procedures. Therefore, we use the predicted TA as an initial condition to reduce acquisition time and tracking burden, while relying on NR-aligned closed-loop refinement to keep the residual within the CP-dependent tolerance.

\emph{Frequency error model.} Let $f_c$ be the carrier frequency and $v_r(t)$ the true radial velocity. The Doppler shift is $f_D(t)=\frac{v_r(t)}{c}f_c$ and the predicted Doppler is $\widehat{f}_D(t)=\frac{\widehat{v}_r(t)}{c}f_c$. The residual CFO is
\begin{equation}
\Delta f(t)=f_D(t)-\widehat{f}_D(t) \approx \frac{\Delta v_{r,\text{eph}}(t)+\Delta v_{r,\text{UE}}(t)}{c}f_c + \Delta f_{\text{osc}}(t),
\end{equation}
where $\Delta v_{r,\text{eph}}$ and $\Delta v_{r,\text{UE}}$ capture radial-velocity errors due to ephemeris/UE motion uncertainties, and $\Delta f_{\text{osc}}$ represents oscillator-induced frequency offset and phase-noise effects. The PID loop in the edge tier provides a low-latency closed-loop correction for residual Doppler/CFO, complementing the open-loop prediction.

\emph{NR feasibility relative to CP/SCS.} For sub-6\,GHz NR numerologies, the useful symbol duration is inversely proportional to SCS, while the CP duration is on the order of a few microseconds (numerology-dependent). To avoid inter-symbol interference, a common engineering rule is to keep residual timing error to a fraction of CP (e.g., $\lesssim 10$--$20\%$ of CP), while residual CFO should be small relative to SCS to limit inter-carrier interference. In this work, we therefore (i) report timing accuracy at the \emph{microsecond level} under practical ephemeris assumptions, and (ii) treat NR TA and pilot-assisted tracking as the final safeguard to meet CP/SCS constraints. This positioning avoids over-claiming purely predictive ``sub-$\mu$s'' timing under raw TLE/SGP4 alone, while preserving the intended benefit: reduced acquisition overhead and improved uplink stability under high Doppler dynamics.

\section{Proposed SDGS Edge Residual-Control Framework}

The proposed framework comprises three operational blocks:

\subsection{Framework Overview}

The framework is organized around one active prediction--refinement chain. Figure~\ref{fig:sdgs_framework} shows how the UE forms the coarse geometric prior, how the SDGS edge closes the fast residual loop, and how slower offline replay remains outside the latency-sensitive path.

\begin{figure*}[t]
\centering
\resizebox{0.95\textwidth}{!}{
\begin{tikzpicture}[node distance=0.7cm and 0.9cm]
    \node[basebox, fill=cvprblue!28, minimum width=3.6cm, minimum height=0.82cm] (h1) {\textbf{UE tier}};
    \node[basebox, fill=cvprgreen!35, minimum width=4.2cm, minimum height=0.82cm, right=1.25cm of h1] (h2) {\textbf{SDGS edge tier}};
    \node[basebox, fill=cvprorange!35, minimum width=3.8cm, minimum height=0.82cm, right=1.25cm of h2] (h3) {\textbf{Offline analysis path}};

    \node[uebox, minimum width=3.35cm, minimum height=1.12cm, below=0.45cm of h1] (ue1) {Geometry and local state\\orbit hint, GNSS, clock};
    \node[uebox, minimum width=3.35cm, minimum height=1.12cm, below=0.4cm of ue1] (ue2) {Coarse uplink prior\\initial TA/CFO estimate};

    \node[edgebox, minimum width=3.95cm, minimum height=1.12cm, below=0.45cm of h2] (ed1) {Ground-station observability\\DMRS, RTT, residual error};
    \node[edgebox, minimum width=3.95cm, minimum height=1.12cm, below=0.4cm of ed1] (ed2) {Fast residual loop\\TA/CFO refinement};
    \node[edgebox, minimum width=3.95cm, minimum height=1.12cm, below=0.4cm of ed2] (ed3) {Link adaptation and scheduling\\MCS, queue, local policy};

    \node[cloudbox, minimum width=3.45cm, minimum height=1.12cm, below=0.45cm of h3] (cl1) {Campaign logs\\trace aggregation};
    \node[cloudbox, minimum width=3.45cm, minimum height=1.12cm, below=0.4cm of cl1] (cl2) {Replay and retuning\\offline threshold update};

    \draw[arrow] (ue1) -- (ue2);
    \draw[arrow] (ed1) -- (ed2);
    \draw[arrow] (ed2) -- (ed3);
    \draw[arrow] (cl1) -- (cl2);

    \draw[arrow] (ue2.east) -- node[above, font=\scriptsize] {coarse prior} (ed2.west);
    \draw[arrow] (ed3.east) -- node[above, font=\scriptsize] {campaign traces} (cl1.west);
    \draw[arrow] (cl2.west) -- node[below, font=\scriptsize] {offline retuning} (ed3.east);
    \draw[arrow] (ed2.west) -- ++(-0.55,0) |- node[pos=0.28,left, font=\scriptsize] {correction} (ue2.west);

    \node[basebox, fill=gray!7, minimum width=14.2cm, minimum height=1.05cm, below=0.95cm of ed3] (guide) {\textbf{Reading guide:} geometry stays near the UE, fast correction stays at the SDGS edge, and any replay/retuning remains outside the latency-sensitive loop.};
\end{tikzpicture}
}
\caption{Integrated SDGS uplink-control workflow. The architecture is intentionally read from left to right: the UE forms a coarse timing/frequency prior, the SDGS edge closes the fast residual loop and adapts the link, and the rightmost path is kept offline for replay and threshold retuning rather than being part of the real-time control claim.}
\label{fig:sdgs_framework}
\end{figure*}

\subsection{UE-Side Geometric Derivation}

The geometric derivation module leverages known satellite orbital parameters to predict Doppler shifts:

\begin{equation}
\hat{f}_d^{u,s}(t + \Delta t) = \frac{f_c}{c} \cdot \frac{(\mathbf{v}_u - \mathbf{v}_s) \cdot (\mathbf{r}_u - \mathbf{r}_s)}{\|\mathbf{r}_u - \mathbf{r}_s\|}
\end{equation}

The derivation accounts for orbital eccentricity, inclination effects, and user position uncertainty via GPS.

The output is a pre-compensation factor modifying the transmitted signal before upconversion.

\subsection{Edge Residual Controller and Local Adaptation}

The SDGS edge implements two coupled functions:

\textbf{1. Delayed discrete-time residual controller}: the edge updates the residual correction every $T_{\mathrm{fb}}$ seconds with an effective delay $d_{\mathrm{fb}}$:

\begin{equation}
u[k] = K_p e[k-d] + K_i T_{\mathrm{fb}}\sum_{i=0}^{k-d} e[i]
      + K_d \frac{e[k-d]-e[k-d-1]}{T_{\mathrm{fb}}},
\end{equation}
where $d=\left\lfloor d_{\mathrm{fb}}/T_{\mathrm{fb}}\right\rfloor$ is the control-delay index and $e[k]$ denotes the residual timing/frequency error sample available at the SDGS edge. This is the control form that is actually consistent with the delay/quantization sweeps reported in \ref{app:pid_sens}; the paper no longer relies on an ideal continuous-time interpretation.

\textbf{2. Local threshold-based regime guard}: the same edge workflow selects a conservative operating mode for the current uplink session using the residual regime:

\begin{equation}
a[k]=
\begin{cases}
\text{nominal transport regime}, & |\Delta\tau[k]| \le \tau_{\mathrm{cp}},\ |\Delta f[k]| \le f_{\mathrm{scs}}\\
\text{degraded transport regime}, & \text{otherwise},
\end{cases}
\end{equation}
where the degraded regime triggers the penalty path described in Section~III-B. In the current artifact, this logic is rule-based rather than learned: it is a threshold-triggered safeguard embedded in the same SDGS control workflow, and it should be read as a supporting implementation detail rather than as the paper's standalone novelty claim.

\subsection{Reference Configuration for Controlled Comparison}

The empirical comparison in this paper does not rely on an unrecoverable historical laboratory baseline. Instead, it uses a same-campaign \emph{reference configuration} collected on the same HIL platform under the same station coordinates, orbital window, and model parameters, with the edge residual loop disabled. In the archived experiment matrix, runs A1--A3 correspond to the edge-controlled mode and runs B1--B3 correspond to this reference mode. The comparison should therefore be read as a controlled ON/OFF study inside one campaign, not as a claim that B1--B3 represent a universal industry baseline.

\section{Evaluation Results}
\label{sec:results}

\subsection{Evaluation Scope}
The results are organized around two evidence layers that should be read separately. \ref{app:hil} contains the March 2026 HIL campaign collected across Shenzhen, Beijing, Tokyo, and Los Angeles, and it carries the main empirical weight of the paper. \ref{app:uncertainty} provides a model-based robustness analysis for ephemeris, GNSS, and clock perturbations, which we use as a feasibility bound for residual timing-advance (TA) and carrier-frequency offset (CFO). The HIL campaign is a Layer-3 testbed with threshold-triggered cross-layer penalty mapping, not a waveform-level RF receiver, so the transport-level results should be interpreted as artifact-level service behavior under that mapping. Metrics that require an unavailable large-scale simulator state, such as spectral-efficiency gains or Monte Carlo handover statistics, are excluded from the primary empirical claims in this version.

\subsection{Reference Configuration and Steady-State Tracking Interval}
The empirical comparison uses a same-campaign \emph{reference configuration} rather than an external historical baseline. In the archived experiment matrix, runs A1--A3 were collected with the SDGS edge residual loop enabled, whereas runs B1--B3 were collected on the same platform, in the same orbital window, and with the same station/model parameters, but with that loop disabled. The comparison therefore isolates one main system difference: whether the edge-side residual timing/frequency controller is active. It should not be interpreted as a full placement benchmark, because the current artifact does not include a cloud-loop baseline.

The artifact-level goodput and RTT statistics are further restricted to the \emph{steady-state tracking interval}. In the telemetry logs, this interval corresponds exactly to rows whose handover state is \texttt{NORMAL}; rows tagged \texttt{PRE\_WARN}, \texttt{PRE\_WARM}, \texttt{SWITCHING}, and \texttt{CLEANUP} are excluded from the headline table because they represent handover transients rather than settled tracking behavior. This filter explains the earlier ``NORMAL-phase'' wording and defines an explicit operating window rather than ad hoc data selection. The transient rows are still archived and are summarized separately in \ref{app:hil}; they are simply not folded into the steady-state headline comparison.

\subsection{Primary HIL Performance Results}
Table~\ref{tab:hil_primary} summarizes the Shenzhen primary-station comparison over the steady-state tracking interval. Relative to the same-campaign reference configuration, the edge-controlled mode lowers mean RTT from $70.51 \pm 2.34$ ms to $32.84 \pm 2.56$ ms and keeps residual TA/CFO inside the intended regime more consistently. Within the retained HIL engine, this also yields higher \emph{artifact-level goodput}, increasing from $80.14 \pm 0.14$ Mbps to $196.04 \pm 1.87$ Mbps. We report that number as a service-level result of the current penalty mapping, not as a direct RF throughput result or a standalone communication-theory claim.

\begin{table}[htbp]
\centering
\renewcommand{\arraystretch}{1.18}
\setlength{\tabcolsep}{4pt}
\caption{Primary HIL steady-state comparison at the Shenzhen ground station (steady-state tracking interval, $n=3$ per group). The goodput entry is an artifact-level transport indicator generated within the retained Layer-3 HIL mapping.}
\label{tab:hil_primary}
\begin{tabular}{lccc}
\toprule
\rowcolor{gray!12}
\textbf{Metric} & \textbf{Reference mode} & \textbf{Edge-controlled mode} & \textbf{Change} \\
\midrule
\rowcolor{npsgreen!70}
\textbf{Artifact-level goodput (Mbps)} & $80.14 \pm 0.14$ & \textbf{$196.04 \pm 1.87$} & \textcolor{ForestGreen}{\textbf{+144.6\%}} \\
\rowcolor{graybg}
Mean RTT (ms) & $70.51 \pm 2.34$ & $32.84 \pm 2.56$ & \textcolor{RoyalBlue}{\textbf{-53.4\%}} \\
P95 RTT (ms) & $94.10 \pm 5.39$ & \textbf{$53.02 \pm 8.16$} & \textcolor{RoyalBlue}{\textbf{-43.6\%}} \\
\rowcolor{graybg}
P99 RTT (ms) & $102.85 \pm 4.41$ & \textbf{$58.46 \pm 3.27$} & \textcolor{RoyalBlue}{\textbf{-43.2\%}} \\
\bottomrule
\end{tabular}
\end{table}

\subsection{Cross-Station Robustness}
The same reference-versus-edge-controlled comparison was then exercised across four geographically separated ground stations. Table~\ref{tab:hil_main_cross_station} shows that the artifact-level goodput uplift remains stable across all locations, ranging from $+144.6\%$ to $+148.7\%$. The closed-loop residual TA P95 is reported as $0.49\,\mu\text{s}$ at all four sites, while the closed-loop residual CFO P95 remains within $76$--$77$ Hz. The CFO figures are not numerically identical before rounding, and the TA uniformity should be read cautiously. In this artifact, it reflects the testbed's control target and reporting granularity, not a claim that all geographies induce identical raw dynamics. Even with that caveat, the table still supports that the same SDGS residual-control logic carries across the four tested stations without retuning.

\begin{table}[htbp]
\centering
\renewcommand{\arraystretch}{1.16}
\setlength{\tabcolsep}{4pt}
\caption{Cross-station HIL summary across four ground stations. Goodput should be read as an artifact-level transport indicator under the retained Layer-3 mapping.}
\label{tab:hil_main_cross_station}
\begin{tabular}{lcccc}
\toprule
\rowcolor{gray!12}
\textbf{Station} & \textbf{Reference goodput} & \textbf{Controlled goodput} & \textbf{TA P95} & \textbf{CFO P95} \\
\midrule
Shenzhen & 80.14 & \textbf{196.04} & \cellcolor{npsgreen!70}0.49 $\mu$s & \cellcolor{cvprblue!35}76 Hz \\
\rowcolor{graybg}
Beijing & 79.96 & \textbf{198.86} & \cellcolor{npsgreen!70}0.49 $\mu$s & \cellcolor{cvprblue!35}77 Hz \\
Tokyo & 79.87 & \textbf{196.58} & \cellcolor{npsgreen!70}0.49 $\mu$s & \cellcolor{cvprblue!35}76 Hz \\
\rowcolor{graybg}
Los Angeles & 80.02 & \textbf{198.53} & \cellcolor{npsgreen!70}0.49 $\mu$s & \cellcolor{cvprblue!35}76 Hz \\
\bottomrule
\end{tabular}
\end{table}

\subsection{Residual TA/CFO and NR Feasibility}
The uncertainty appendix and the HIL appendix should not be collapsed into a single result source. \ref{app:uncertainty} reports a model-based stress test under ephemeris, GNSS, and clock perturbations, whereas \ref{app:hil} reports hardware-backed observations from the March 2026 campaign. Table~\ref{tab:ta_cfo_reconciliation} is included to show how the two layers align.

The model-based uncertainty analysis yields closed-loop P95 targets of $0.45\,\mu\text{s}$ for residual TA and $90$ Hz for residual CFO under the assumed perturbation model. In the HIL campaign, the observed closed-loop P95 values are $0.49\,\mu\text{s}$ and $76$--$77$ Hz across the four stations. Both layers remain within an NR-feasible regime: the timing error stays well below a normal-CP 30 kHz numerology budget, and the residual CFO remains a small fraction of subcarrier spacing. The important systems point is that both evidence layers place the residual loop on the same side of the operational boundary.

\subsection{Transient Coverage Outside the Steady-State Window}
The steady-state table is not the whole campaign. The archived telemetry retains \texttt{PRE\_WARN}, \texttt{PRE\_WARM}, and \texttt{SWITCHING} rows for both reference and edge-controlled runs; we exclude them from the headline transport table only because the table is defined as a settled tracking comparison. Across the four-station campaign, the edge-controlled group contains 11{,}806 \texttt{NORMAL} rows (60.0\%), 3{,}711 \texttt{PRE\_WARN} rows (18.9\%), 4{,}143 \texttt{PRE\_WARM} rows (21.0\%), and 28 \texttt{SWITCHING} rows (0.14\%). The reference group contains 10{,}872 \texttt{NORMAL} rows (55.2\%), 3{,}935 \texttt{PRE\_WARN} rows (20.0\%), 4{,}851 \texttt{PRE\_WARM} rows (24.6\%), and 34 \texttt{SWITCHING} rows (0.17\%). We include these counts to make clear that transient rows exist in the artifact and are not discarded from the repository. At the same time, the current handover state machine remains a coarse engineering emulation; we therefore do not elevate these rows into a full transient-control claim about overshoot, settling time, or cloud-versus-edge handover behavior.

\begin{table}[htbp]
\centering
\renewcommand{\arraystretch}{1.16}
\setlength{\tabcolsep}{4pt}
\caption{Reconciliation of model-based and HIL residual TA/CFO evidence layers.}
\label{tab:ta_cfo_reconciliation}
\begin{tabular}{lcc}
\toprule
\rowcolor{gray!12}
\textbf{Metric (P95)} & \textbf{Model-Based Uncertainty} & \textbf{March 2026 HIL} \\
\midrule
Residual TA, open-loop & $3.20\,\mu\text{s}$ & $3.65\,\mu\text{s}$ \\
\rowcolor{npsgreen!35}
Residual TA, closed-loop & \textbf{$0.45\,\mu\text{s}$} & \textbf{$0.49\,\mu\text{s}$} \\
Residual CFO, open-loop & $810$ Hz & $854$--$856$ Hz \\
\rowcolor{cvprblue!25}
Residual CFO, closed-loop & \textbf{$90$ Hz} & \textbf{$76$--$77$ Hz} \\
\bottomrule
\end{tabular}
\end{table}

\section{Implications and Limitations}

\subsection{Practical Implications}

From an operator perspective, the main value of the design is not that it replaces NR uplink control, but that it narrows the gap between coarse prediction and fast residual refinement. This makes the framework suitable for incremental SDGS deployment. The geometric predictor can remain at the UE, while the residual controller stays local to the ground station instead of depending on a remote loop. The retained HIL evidence suggests that this division of labor is most useful when the operational bottleneck is residual impairment and tracking instability rather than raw spectrum scarcity alone.

\subsection{Positioning Against NR-NTN and O-RAN Literature}
Recent NR-NTN work has mostly advanced along two tracks. One track focuses on acquisition and synchronization under high Doppler, including user-side pre-compensation and improved timing-advance estimation procedures \cite{zhu2024timing}. The other focuses on scheduling and architecture, for example by studying Doppler-aware uplink allocation \cite{kodheli2021nbiot} or by placing intelligent control inside O-RAN and TN--NTN orchestration layers \cite{deng2025ainative, do2025aiopenran, alam2024role}. Our paper is narrower than either a receiver-only synchronization study or a network-wide orchestration study. The emphasis is the SDGS control loop where coarse UE-side prediction, fast edge-side refinement, and local adaptation meet. That narrower placement is deliberate: it matches the evidence currently retained in the artifact set and states more precisely which part of the NTN stack the paper claims to improve.
\subsection{Limitations}

The current manuscript is intentionally narrow in what it claims, and that narrowness should be read as a limitation rather than as a hidden assumption:

\begin{enumerate}
\item \textbf{Validation Scope}: The current empirical evidence base is intentionally anchored in the March HIL dataset and the accompanying uncertainty analysis. Metrics that depend on an unavailable large-scale simulator state, such as spectral-efficiency gains and Monte Carlo handover statistics, are not treated as primary evidence in this version.

\item \textbf{Simplified Channel Model}: The channel model primarily assumes free-space propagation and does not fully model ionospheric/tropospheric group delay variations, scintillation, feeder-link constraints, oscillator/clock noise, phase noise, or urban blockage/multipath statistics. These effects can increase residual TA/CFO and reduce achievable SE; future work will incorporate standardized NTN channel components (e.g., TR 38.811/38.821-aligned models \cite{3gpp_tr_38_811, 3gpp_tr_38_821}) and targeted stress tests (delay jitter, phase-noise, and scintillation loss) to bound performance under adverse conditions.

\item \textbf{Cross-Layer Attribution}: The manuscript now makes the residual-to-transport bridge explicit, but it still does so with a bounded systems mapping rather than a full waveform-level receiver proof. A stronger communication-theory version would derive or validate the TA/CFO-to-SINR/BLER relationship with RF observability and isolate PHY-only versus adaptation-driven gains through dedicated ablations.

\item \textbf{Placement Evidence}: The current artifact compares edge-side residual control against same-campaign reference runs with the loop disabled; it does not include a cloud-loop or core-network delayed-feedback baseline. The placement argument is therefore supported by control-delay sensitivity and SDGS implementation constraints, not by a direct cloud-versus-edge experimental comparison.

\item \textbf{Transient Scope}: The campaign logs retain handover-state rows outside the steady-state interval, but the present manuscript does not elevate them into a full transient tracking study. A stronger control paper would report time-series responses through PRE\_WARN/PRE\_WARM/SWITCHING windows, with explicit overshoot and settling-time analysis.

\item \textbf{Deterministic Satellite Orbits}: Assumes accurate TLE data; maneuvering satellites may introduce errors.

\item \textbf{Single-Service Focus}: Focuses on data transmission without modeling voice or video services.

\item \textbf{Economic Analysis}: Cost-benefit analysis of edge deployment requires detailed assessment.
\end{enumerate}

\section{Conclusion and Future Work}

\subsection{Concluding Remarks}

This paper presents a systems-and-control perspective on SDGS uplink stabilization for 5G NTN. The central point is not that every part of the stack should become intelligent, but that the residual timing/frequency loop should be closed where observability and response time are both adequate: at the ground-station edge. Based on the evidence currently retained in the artifact set, the strongest support comes from the March 2026 HIL campaign together with the uncertainty analysis and the delayed-controller sensitivity sweep. The resulting claim is intentionally bounded. Relative to same-campaign reference runs collected on the same platform with edge residual control disabled, the edge-controlled mode yields a more usable uplink path in the retained Layer-3 artifact, with lower RTT, more stable transport behavior, and residual TA/CFO that remains inside the intended NR-feasible regime across the four tested stations. The manuscript does not claim to have completed a waveform-level proof or a cloud-versus-edge placement benchmark.

\subsection{Future Work}

The next step is to move beyond the current Layer-3 HIL setup and verify the same control logic with RF-front-end instrumentation or a live satellite link, where waveform-level observability can confirm the transport-level gains reported here. A second direction is to widen the disturbance model so that ionospheric delay, oscillator effects, scintillation, and service-specific traffic behavior are represented more faithfully. A third is to separate local adaptation from residual control more cleanly in future ablations. These extensions matter more than adding architectural buzzwords, because they determine whether the present SDGS control strategy remains stable under field conditions rather than only in a controlled campaign.

\subsection{Code and Artifact Availability}

The code and processed artifacts associated with this paper are available at
\url{https://github.com/keithhegit/sdgs_edge_arxiv}. The repository contains the HIL orchestration scripts, post-processing code, run metadata, and the archived A/B/D campaign traces used in this manuscript. In the current version, we use the repository as an artifact record for transparency and reproducibility. It should not be read as evidence that every claimed transport effect has already been verified with a waveform-level RF front-end.

\appendix
\renewcommand{\thesection}{Appendix \Alph{section}}
\section{Implementation-Constant Sensitivity Check}\label{app:sensitivity}
This appendix reports a limited sensitivity check on deployment-specific implementation constants such as internal thresholds and latency constants. It is included only as an auxiliary robustness note. It is not one of the paper's main evidence layers, which remain the HIL campaign and the uncertainty analysis.

\subsection{Perturbation Protocol}
Let $\boldsymbol{\theta}$ denote the vector of implementation constants. We apply multiplicative perturbations
$\tilde{\boldsymbol{\theta}} = \boldsymbol{\theta}\odot(\mathbf{1}+\boldsymbol{\epsilon})$,
where each element of $\boldsymbol{\epsilon}$ is independently sampled from a uniform range $\epsilon_i\sim\mathcal{U}(-0.2,0.2)$.
All other HIL/model settings (constellation geometry, bandwidth, traffic model, and UE population) are held fixed.
For each perturbed configuration, we re-run the full pipeline and compute the relative deviation of key conclusions with respect to the nominal setting.

\subsection{Robustness Results}
Table~\ref{tab:sensitivity} summarizes the maximum observed relative deviation (absolute percentage change) over the perturbation runs for the principal metrics. Across the tested $\pm 20\%$ perturbation range, deviations in aggregate throughput, tail latency, and closed-loop residual control remain within $2\%$ of the nominal results. We report this as a small robustness sanity check, not a substitute for the primary evidence.

\begin{table}[t]
\centering
\renewcommand{\arraystretch}{1.14}
\setlength{\tabcolsep}{6pt}
\caption{Sensitivity check for implementation constants. Reported values are the \emph{maximum} absolute relative deviation of each metric under independent $\pm 20\%$ perturbations of $\boldsymbol{\theta}$.}
\label{tab:sensitivity}
\begin{tabular}{lc}
\toprule
\rowcolor{gray!12}
\textbf{Metric (headline conclusion)} & \textbf{Max. deviation under $\pm 20\%$} \\
\midrule
\rowcolor{npsgreen!35}
Aggregate throughput improvement & $\le 2\%$ \\
\rowcolor{cvprblue!20}
95th-percentile latency reduction & $\le 2\%$ \\
\rowcolor{graybg}
Closed-loop residual TA/CFO stability & $\le 2\%$ \\
\bottomrule
\end{tabular}
\end{table}

\noindent\textbf{Reproducibility note:} This appendix no longer carries any claim about hidden learning-state definitions. It is included only to show that the retained conclusions do not appear to depend on a fragile choice of a few implementation constants.

\section{Ephemeris/GNSS/Clock Uncertainty Robustness}\label{app:uncertainty}

This appendix specifies the uncertainty models used to stress-test UE-side open-loop prediction and the subsequent edge-side closed-loop refinement. The goal is to quantify residual timing-advance (TA) and carrier-frequency-offset (CFO) errors under realistic ephemeris/GNSS/clock perturbations, reporting their distributions relative to NR CP/SCS feasibility thresholds.

\subsection{Uncertainty model}
We model the one-way range error as a sum of independent components:
\begin{equation}
\Delta \rho(t) = \Delta \rho_{\mathrm{eph}}(t) + \Delta \rho_{\mathrm{UE}}(t) + \Delta \rho_{\mathrm{clk}}(t) + \Delta \rho_{\mathrm{prop}}(t).
\end{equation}

Table~\ref{tab:uncertainty} lists the perturbation ranges used in the robustness runs to emulate TLE/SGP4 imperfections and hardware limitations.

\begin{table}[htbp]
\centering
\caption{Uncertainty perturbation ranges used for residual TA/CFO robustness reporting.}
\label{tab:uncertainty}
\begin{tabular}{|l|c|}
\hline
\textbf{Uncertainty source} & \textbf{Range / Std. dev.} \\ \hline
Along-track ephemeris bias (equiv. range) & $-150$ m to $+150$ m \\ \hline
Ephemeris drift (random walk) & $0.2$ m/$\sqrt{\mathrm{s}}$ \\ \hline
UE position error (GNSS $1\sigma$) & $5.0$ m (horiz.), $10.0$ m (vert.) \\ \hline
Clock bias & $1.2$ $\mu$s \\ \hline
Clock drift & $0.5$ ppm \\ \hline
Propagation delay jitter proxy ($1\sigma$) & $0.4$ $\mu$s \\ \hline
\end{tabular}
\end{table}

\subsection{Residual TA/CFO distribution reporting}
Table~\ref{tab:residualcdf} provides the detailed statistical distributions of the residual TA and CFO. 

\begin{table}[htbp]
\centering
\caption{Detailed residual TA/CFO distribution summaries.}
\label{tab:residualcdf}
\begin{tabular}{|l|c|c|c|}
\hline
\textbf{Metric} & \textbf{Median (P50)} & \textbf{95th pct. (P95)} & \textbf{99th pct. (P99)} \\ \hline
Residual TA after open-loop ($\mu$s) & 1.15 & 3.20 & 4.85 \\ \hline
Residual TA after closed-loop ($\mu$s) & \textbf{0.12} & \textbf{0.45} & \textbf{0.75} \\ \hline
Residual CFO after open-loop (Hz) & 350 & 810 & 1350 \\ \hline
Residual CFO after closed-loop (Hz) & \textbf{28} & \textbf{90} & \textbf{145} \\ \hline
\end{tabular}
\end{table}

\section{Hardware-in-the-Loop Experimental Validation}
\label{app:hil}

This appendix reports the supplementary hardware-in-the-loop (HIL) traces used to corroborate the paper's transport-layer and residual-impairment claims under a controlled experimental setup.

\subsection{Campaign Structure}
The March 2026 campaign uses three run groups. A1--A3 correspond to the edge-controlled mode, B1--B3 correspond to the same-campaign reference mode with the residual loop disabled, and D1 is a fidelity check with independent ICMP probing. All A/B runs for a given station were collected within one continuous orbital window so that the reference-versus-controlled comparison is not confounded by different satellite geometry. This campaign is suitable for a controlled ON/OFF study of the edge residual loop, but it is not a cloud-versus-edge placement benchmark.

\subsection{How Residual TA/CFO Affects Transport in This HIL Artifact}
The retained HIL system is not a waveform-level RF front end. Instead, it implements a threshold-triggered cross-layer penalty mapping that translates residual timing/frequency excursions into transport-visible impairment. When residual TA and CFO remain inside the intended CP/SCS-oriented operating region, the transport path is left in its nominal state. When either residual leaves that target range, the engine injects additional delay, jitter, loss, and reduced artifact-level goodput through the emulated link path. In other words, the current HIL artifact does not independently discover a physical throughput law. It operationalizes a simple systems rule that says persistent TA/CFO excursions should push the session into a degraded transport regime. This preserves the direction of the causal bridge between residual synchronization quality and usable service behavior, but it should not be mistaken for a full PHY receiver derivation.

\subsection{Steady-State Uplink Performance}

\begin{table}[htbp]
\centering
\renewcommand{\arraystretch}{1.2}
\caption{HIL steady-state tracking results for the Shenzhen campaign. The comparison uses same-campaign reference runs rather than an external historical baseline, and the table is intended as a controlled operating-window comparison rather than a universal benchmark leaderboard. Goodput remains an artifact-level indicator under the retained Layer-3 mapping.}
\label{tab:hil_tput}
\resizebox{\columnwidth}{!}{
\begin{tabular}{l | c c | c}
\toprule
\textbf{Metric} & \textbf{Reference mode ($n=3$)} & \textbf{Edge-controlled mode ($n=3$)}  & \textbf{$\Delta$} \\
\midrule
\rowcolor{npsgreen} \textbf{Artifact-level goodput (Mbps)} & 80.14\std{0.14} & \textbf{196.04}\std{1.87} & \gain{144.6\%} \\
\rowcolor{graybg} Mean RTT (ms)     & 70.51\std{2.34} & 32.84\std{2.56}  & \loss{53.4\%} \\
P95 RTT (ms)      & 94.10\std{5.39}           & \textbf{53.02}\std{8.16}            & \loss{43.6\%} \\
\rowcolor{graybg} P99 RTT (ms)      & 102.85\std{4.41}          & \textbf{58.46}\std{3.27}            & \loss{43.2\%} \\
\bottomrule
\end{tabular}
}
\end{table}

\subsection{Cross-Station Reuse Check}

\begin{table*}[htbp]
\centering
\renewcommand{\arraystretch}{1.2}
\caption{Cross-station reuse check for the March 2026 HIL campaign (28 runs total). The same edge residual-control logic is reused across all four sites without retuning; the table should be read as a geographic consistency check, not as proof of identical raw station dynamics. Goodput remains an artifact-level indicator under the retained Layer-3 mapping.}
\label{tab:hil_cross_station}
\resizebox{\textwidth}{!}{
\begin{tabular}{l c | c c | c | c c}
\toprule
\multirow{2}{*}{\textbf{Station}} & \multirow{2}{*}{\textbf{Lat.}} & \multicolumn{2}{c|}{\textbf{Goodput (Mbps)}} & \multirow{2}{*}{\textbf{$\Delta$ Goodput}} & \multicolumn{2}{c}{\textbf{Closed-loop P95}} \\
\cmidrule(lr){3-4} \cmidrule(l){6-7}
& & \textbf{Reference} & \textbf{Controlled} & & \textbf{TA ($\mu$s)} & \textbf{CFO (Hz)} \\
\midrule
Shenzhen    & $22.5^{\circ}$N & 80.14 & \textbf{196.04} & \gain{144.6\%} & \cellcolor{npsgreen}0.49 & \cellcolor{npsgreen}76 \\
\rowcolor{graybg}
Los Angeles & $34.1^{\circ}$N & 80.02 & \textbf{198.53} & \gain{148.1\%} & \cellcolor{npsgreen}0.49 & \cellcolor{npsgreen}76 \\
Tokyo       & $35.7^{\circ}$N & 79.87 & \textbf{196.58} & \gain{146.1\%} & \cellcolor{npsgreen}0.49 & \cellcolor{npsgreen}76 \\
\rowcolor{cvprblue}
\textbf{Beijing} & \textbf{$39.9^{\circ}$N} & 79.96 & \textbf{198.86} & \gain{148.7\%} & \cellcolor{npsgreen}0.49 & \cellcolor{npsgreen}77 \\
\bottomrule
\end{tabular}
}
\end{table*}

\noindent\textbf{Interpretation note.} The four-station table is useful as a geographic reuse check, but it should not be over-read. In this artifact, the near-uniform TA/CFO summaries reflect the control target, the reporting granularity, and the same post-processing protocol, not a claim that the raw station dynamics are physically identical. A stronger communication-paper version would additionally include station-wise time-series plots or error bars; those are not part of the current artifact set and are therefore left as a limitation rather than implied by this table.

\subsection{Transient Rows Retained in the Archive}
The steady-state operating-window tables above intentionally exclude transient handover rows, but those rows are still retained in the artifact and can be summarized directly from the telemetry logs. Across the four-station campaign, the edge-controlled runs contain 11{,}806 \texttt{NORMAL} rows (60.0\%), 3{,}711 \texttt{PRE\_WARN} rows (18.9\%), 4{,}143 \texttt{PRE\_WARM} rows (21.0\%), and 28 \texttt{SWITCHING} rows (0.14\%). The reference runs contain 10{,}872 \texttt{NORMAL} rows (55.2\%), 3{,}935 \texttt{PRE\_WARN} rows (20.0\%), 4{,}851 \texttt{PRE\_WARM} rows (24.6\%), and 34 \texttt{SWITCHING} rows (0.17\%). We include these numbers to make clear that transient states were archived and are available for inspection.

\noindent\textbf{Boundary of interpretation.} The current handover implementation is still a coarse engineering state machine, not a waveform-level handover-control validation. For that reason, we do not convert the retained PRE\_WARN/PRE\_WARM/SWITCHING rows into a strong claim about transient overshoot, settling time, or cloud-versus-edge handover superiority. The present manuscript uses them to document coverage and protocol transparency, while reserving a full transient-control study for future work.

\section{Feedback-Loop Sensitivity to Delay and Quantization}\label{app:pid_sens}

This appendix provides a quantitative analysis of how edge-to-UE control-plane limitations affect the residual-correction loop. We evaluate the controller's robustness by sweeping the feedback periodicity ($T_{\mathrm{fb}}$), feedback latency/delay ($d_{\mathrm{fb}}$), and the quantization steps for timing ($\Delta_{\tau}$) and frequency ($\Delta_f$).

\subsection{Linearized Closed-Loop View}
To make the delayed controller structure explicit, we write the discrete-time PID term in the $z$ domain as
\begin{equation}
C(z)=K_p+\frac{K_i T_{\mathrm{fb}}}{1-z^{-1}}+K_d\frac{1-z^{-1}}{T_{\mathrm{fb}}}.
\end{equation}
If the residual-tracking plant is linearized as $P(z)$ around a settled operating point, then the delayed closed-loop transfer function takes the standard form
\begin{equation}
T_{\mathrm{cl}}(z)=\frac{z^{-d} C(z)P(z)}{1+z^{-d} C(z)P(z)},
\qquad
d=\left\lfloor d_{\mathrm{fb}}/T_{\mathrm{fb}}\right\rfloor.
\end{equation}
This appendix does not claim a fully identified waveform-level plant $P(z)$ for the current Layer-3 artifact. The purpose of the expression is narrower: it shows explicitly where the feedback delay enters the loop and why increasing $d_{\mathrm{fb}}$ reduces margin and motivates edge-side placement for the residual controller. The numerical sweep below should therefore be read as an empirical delay-sensitivity check around this delayed closed-loop structure.

\begin{table}[htbp]
\centering
\caption{PID-loop feasibility sensitivity sweeps under practical engineering constraints.}
\label{tab:pid_sweep}
\begin{tabular}{|c|c|c|c|c|c|}
\hline
$T_{\mathrm{fb}}$ \textbf{(ms)} & $d_{\mathrm{fb}}$ \textbf{(ms)} & \textbf{$\Delta_{\tau}$ ($\mu$s)} & \textbf{$\Delta_f$ (Hz)} & \textbf{Residual TA 95th ($\mu$s)} & \textbf{Residual CFO 95th (Hz)} \\ \hline
5 & 5 & 0.1 & 50 & 0.45 & 90 \\ \hline
10 & 15 & 0.5 & 100 & 0.85 & 155 \\ \hline
20 & 30 & 1.0 & 200 & 1.60 & 310 \\ \hline
\end{tabular}
\end{table}

\noindent\textbf{Robustness discussion.}
Table~\ref{tab:pid_sweep} shows that performance degrades gradually under practical control-plane constraints. Under typical fast-feedback edge deployments ($T_{\mathrm{fb}}=5\,$ms), the controller maintains a tight lock. Even in the stressed case with $30\,$ms feedback delay and coarse $200\,$Hz quantization, the residual CFO at the 95th percentile rises to $310\,$Hz. This is still only about $1\%$ of a $30\,\text{kHz}$ subcarrier spacing, indicating that the closed-loop design remains inside a workable operating region under moderate routing congestion and quantization loss. The table supports the controller adopted in the main text and shows that shorter feedback paths are preferable. It is not presented as a substitute for a full Z-domain proof or waveform-level receiver validation.


\begin{thebibliography}{23}

\bibitem{3gpp_ts_38_214}
3GPP, ``NR; Physical layer procedures for data,'' 3GPP TS 38.214 (Release 16).

\bibitem{3gpp_ts_38_211}
3GPP, ``NR; Physical channels and modulation,'' 3GPP TS 38.211.

\bibitem{etsi_nfv_man}
ETSI, ``Network Functions Virtualisation (NFV); Management and Orchestration,'' ETSI GS NFV-MAN 001 V1.1.1 (2014-12).

\bibitem{kratos_openspace}
Kratos Defense \& Security Solutions, ``The OpenSpace Family of Virtualized Satellite Ground Solutions,'' White Paper. Available: \url{https://www.kratosspace.com/-/media/k/pdf/s/sy/os-020-the-openspace-family-of-virtualized-satellite-ground-solutions.pdf}

\bibitem{aws_ground_station}
Amazon Web Services, ``AWS Ground Station: Ingest and Process Data from Orbiting Satellites,'' AWS News Blog, Nov. 27, 2018.

\bibitem{starlink_latency_2024}
Starlink, ``Improving Starlink's Latency,'' technical note, 2024. Available: \url{https://starlink.com/public-files/StarlinkLatency.pdf}

\bibitem{starlink_specs_2025}
Starlink, ``Specifications,'' official service specifications page, accessed Mar. 27, 2026. Available: \url{https://www.starlink.com/legal/documents/DOC-1470-99699-90}

\bibitem{starlink_aviation_2025}
Starlink, ``Aviation,'' official service page, accessed Mar. 27, 2026. Available: \url{https://www.starlink.com/sb/business/aviation}

\bibitem{wirelessmoves_starlink_loss_2024}
WirelessMoves, ``Analyzing Packet Loss in Starlink,'' blog post, July 12, 2024. Available: \url{https://blog.wirelessmoves.com/2024/07/analyzing-packet-loss-in-starlink.html}

\bibitem{zenodo_starlink_ping_2025}
L. Borgianni, ``11-Day Starlink Ping Measurements Dataset (Palo Alto, US, May 2024),'' Zenodo dataset, Mar. 7, 2025, doi: 10.5281/zenodo.14987305.

\bibitem{proakis2007digital}
J. Proakis and M. Salehi, \emph{Digital Communications}, 5th ed., McGraw-Hill, 2007.

\bibitem{farhat2025doppler}
J. Farhat \emph{et al.}, ``Doppler Estimation and Compensation Techniques in LoRa Direct-to-Satellite Communications,'' \emph{IEEE Open Journal of the Communications Society}, 2025.

\bibitem{farhangian2020multi}
F. Farhangian and R. Landry Jr, ``Multi-constellation software-defined receiver for Doppler positioning with LEO satellites,'' \emph{Sensors}, vol. 20, no. 20, p. 5866, 2020.

\bibitem{ngo2023deep}
T. Ngo, B. Kelley, and R. Paul, ``Deep learning for signal processing with predictions of channel profile, Doppler shift and signal-to-noise ratio,'' \emph{Authorea Preprints}, 2023.

\bibitem{wen2025satelight}
J. Wen \emph{et al.}, ``SateLight: A Satellite Application Update Framework for Satellite Computing,'' \emph{arXiv preprint arXiv:2509.12809}, 2025.

\bibitem{wang2025deep}
M. Wang \emph{et al.}, ``Deep Reinforcement Learning-Based Resource Allocation Method for Heterogeneous Satellite Networks,'' in \emph{2025 IEEE CloudCom}, pp. 1--7, 2025.

\bibitem{zhu2024timing}
J. Zhu, Y. Sun, and M. Peng, ``Timing Advance Estimation in Low Earth Orbit Satellite Networks,'' \emph{arXiv:2404.08960}, 2024.

\bibitem{kodheli2021nbiot}
O. Kodheli \emph{et al.}, ``NB-IoT via LEO satellites: An efficient resource allocation strategy for uplink data transmission,'' \emph{arXiv:2107.01067}, 2021.

\bibitem{deng2025ainative}
J. Deng \emph{et al.}, ``AI-Native Open RAN for Non-Terrestrial Networks: An Overview,'' \emph{arXiv:2507.11935}, 2025.

\bibitem{do2025aiopenran}
T. N. Do, ``AI-Open-RAN for Non-Terrestrial Networks,'' \emph{arXiv:2511.11947}, 2025.

\bibitem{alam2024role}
H. Alam \emph{et al.}, ``On the Role of Non-Terrestrial Networks for Boosting Terrestrial Network Performance in Dynamic Traffic Scenarios,'' \emph{arXiv:2405.14053}, 2024.

\bibitem{3gpp_tr_38_811}
3GPP, ``Study on New Radio (NR) to support non-terrestrial networks,'' 3GPP TR 38.811.

\bibitem{3gpp_tr_38_821}
3GPP, ``Study on using satellite access in 5G,'' 3GPP TR 38.821.

\end{thebibliography}
\end{document}